\documentclass[aps,prb,twocolumn,showpacs]{revtex4}
\usepackage{graphicx}

\begin{document}
\title{Frustrated two-dimensional Josephson junction array near incommensurability}
\author{In-Cheol Baek}
\author{Young-Je Yun}
\author{Mu-Yong Choi}
\affiliation{BK21 Physics Division and Institute of Basic Science, Sungkyunkwan University,
Suwon 440-746, Korea}

\begin{abstract}
To study the properties of frustrated two-dimensional Josephson junction arrays near incommensurability, we examine the current-voltage characteristics of a square proximity-coupled Josephson junction array at a sequence of frustrations $f=$ 3/8, 8/21, 0.382 $(\approx (3-\sqrt{5})/2)$, 2/5, and 5/12.
Detailed scaling analyses of the current-voltage characteristics reveal approximately universal scaling behaviors for $f=$ 3/8, 8/21, 0.382, and 2/5.
The approximately universal scaling behaviors and high superconducting transition temperatures indicate that both the nature of the superconducting transition and the vortex configuration near the transition at the high-order rational frustrations $f=$ 3/8, 8/21, and 0.382 are similar to those at the nearby simple frustration $f=$ 2/5.
This finding suggests that the behaviors of Josephson junction arrays in the wide range of frustrations might be understood from those of a few simple rational frustrations.
\end{abstract}

\pacs{74.81.Fa, 64.60.Cn, 64.70.Pf, 74.25.Qt}

\maketitle

For Josephson junction arrays (JJA's), frustration can be introduced by applying an external magnetic field.
In the presence of frustration, a finite density of vortices is induced in the array. 
For a two-dimensional (2D) JJA, the mean number of vortices per plaquette induced by a uniform external field is identical with the frustration index $f$ defined as the fraction of a flux quantum per plaquette. 
The competition between two different length scales, the mean separation of vortices and the period of the underlying pinning potential of the array, leads to various structures of vortex lattices at low temperatures.
At a simple rational $f$, the ground state is a commensurate pinned vortex lattice.\cite{R1}
For example, vortices in a square array at $f=$ 1/2, 1/3, or 2/5 have a staircase state with a quasi-one-dimensional structure in a diagonal direction.\cite{R2}
For $f=$ 1/3 or 2/5, the high-temperature isotropic vortex liquid freezes into the pinned vortex lattice with the staircase configuration through a finite-temperature superconducting transition.
In the weak magnetic-field limit, the superconducting vortex lattice becomes triangular.\cite{R3}
In a high-order rational or an irrational field, however, both the low-temperature vortex configuration and the nature of the superconducting transition remain controversial.
Some years ago, Halsey \cite{R4} suggested, based on Monte Carlo (MC) studies of the $XY$ model, that in the limit of the irrational $f^{*}= (3 - \sqrt{5})/2$, there may exist a glass transition to a superconducting disordered vortex state at some finite temperature $T_g \approx$ 0.25$J/k_{B}$ for a square array, in which $J$ is the junction coupling strength.
Other arguments,\cite{R5,R6,R7} however, suggest that the glass transition appears at zero temperature and that the superconducting transitions at rational $f$'s near $f^*$ should occur at some finite temperatures which decrease to zero as $f$ approaches the irrational value of $f^*$.
Some of the recent MC simulations,\cite{R8} on the other hand, showed that at most $f$'s in the range 1/3 $\le f \le$ 1/2 except a few simple rational $f$'s, vortices in a square JJA undergo two separate transitions at finite temperatures: a sharp first-order phase transition at temperature $T_o=$ (0.15$-$0.2)$J/k_{B}$ to a resistive quasiordered state and a pinning transition at a lower temperature to a superconducting ordered state with the vortex configuration consisting of periodic arrangements of either staircase patterns of the simple rational $f$'s or hole defects on the $f=$ 1/2 configuration.
A single finite-temperature phase transition from a vortex-liquid phase to a superconducting ordered phase consisting of domains separated by parallel domain walls has also been suggested by a different numerical work for a sequence of high-order rational $f$'s approaching the irrational $f^*$.\cite{R9}
The nature of the superconducting transition and the vortex structure of the superconducting ground state near incommensurability obtained by numerical efforts have been found \cite{R8,R9} to strongly depend upon the imposed boundary condition and choice of dynamics.

There have been very few experimental studies about phase transitions near incommensurability. 
They agree on a single finite-temperature phase transition to a superconducting state which can be understood as a pinned vortex phase with long-range phase coherence.\cite{R10,R11}
However, since the experiments were performed on superconducting wire networks where phase fluctuations were much weaker than in JJA's, the superconducting behavior is suspected by some to be dominated by a mean-field transition.\cite{R5}
Besides, the experiments provided only limited information about the vortex structure of the superconducting state.
In addition to stronger phase fluctuations, a proximity-coupled JJA has a broader critical region than a superconducting wire network, which enables one to investigate the superconducting transition with higher precision.
For a JJA, it is also possible to measure the single-junction critical current $i_c$ and determine the phase-transition temperature in units of $J/k_B$ ($= \hbar i_c / 2ek_B$).
Knowing the phase-transition temperature in units of $J/k_B$, one can make a direct comparison between experimental results and the numerical expectations. 
In this paper, we present an experimental investigation of the superconducting phase transition of a proximity-coupled JJA near incommensurability, that is, at high-order rational frustrations near the irrational $f^{*} = (3-\sqrt{5})/2$.
We examine the superconducting scaling behavior of the current-voltage ($IV$) characteristics of a square array at $f=$ 3/8, 8/21, 0.382 ($\approx (3 - \sqrt{5})/2$), 2/5, and 5/12.
Detailed scaling analyses reveal approximately universal scaling behaviors for $f=$ 3/8, 8/21, 0.382, and 2/5.
The superconducting transition temperatures are found as high as $(0.19-0.22) J/k_B$. 
We discuss implications of the universal scaling behavior and the high superconducting transition temperatures in conjunction with the nature of the superconducting transition and the vortex configuration of the superconducting state near incommensurability.

The experiments were performed on a square array of 200$\times$1000 Nb/Cu/Nb Josephson junctions described in Ref. 12. 
The array completed the zero-field superconducting transition at $T \approx$ 6.24 K with a transition width $\approx$ 0.4 K.
The appearance of many resistance minima in the magnetoresistance curve confirms the good uniformity of the sample.
The frustration could be precisely adjusted by the use of the magnetoresistance curve of the sample showing sharp resistance minima at fractional $f$'s.\cite{R13}
The standard four-probe method utilizing a transformer-coupled lock-in voltmeter and a square-wave current at 23 Hz was used for the $IV$ characteristics measurements. 
The single-junction critical current $i_c$ and the junction coupling strength $J(=\hbar i_c / 2e)$ at high temperatures were determined by extrapolating the $i_c$ vs $T$ data at low temperatures by the use of de Gennes expression \cite{R14} in the dirty limit.
The $i_c$ at low temperatures was obtained from the $I$ vs $dV/dI$ curves.

\begin{figure}[t]
\includegraphics[width=1.0\linewidth]{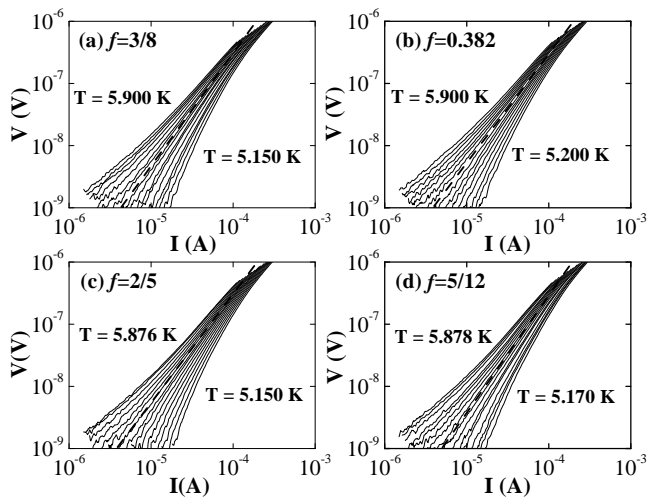}
\caption{Some of $IV$ curves for four different frustrations: (a) $f=$ 3/8 at $T=$ 5.150 to 5.900 K, (b) $f=$ 0.382 at $T=$ 5.200 to 5.900 K, (c) $f=$ 2/5 at $T=$ 5.150 to 5.876 K, and (d) $f=$ 5/12 at $T=$ 5.170 to 5.878 K. The dashed lines are drawn to show the power law ($V \sim I^{z+1}$) behavior at the critical temperature.}
\label{fig1}
\end{figure}

\begin{figure}[t]
\includegraphics[width=1.0\linewidth]{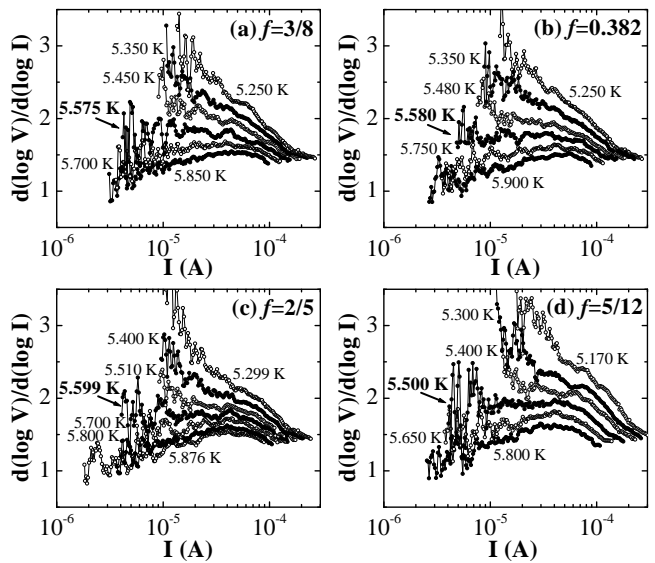}
\caption{Current dependences of the slope of the $IV$ curves in Fig.\ \ref{fig1}.}
\label{fig2}
\end{figure}

\begin{figure}[t]
\includegraphics[width=1.0\linewidth]{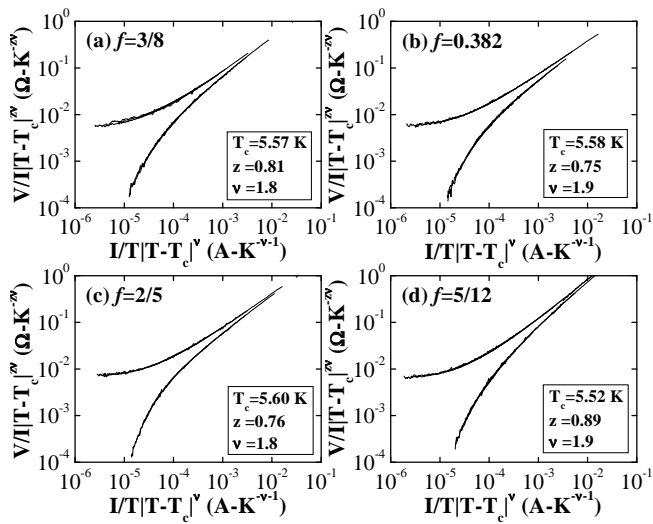}
\caption{Scaling plots of the $IV$ curves. Each plot contains $IV$ curves at 18$-$21 different temperatures. The values of $T_c$, $z$, and $\nu$ used to scale the data are shown in the insets.}
\label{fig3}
\end{figure}

The $IV$ characteristics of the sample for $f=$ 3/8, 0.382, 2/5, and 5/12 are shown in Fig.\ \ref{fig1}.
Relatively narrow voltage ranges of the $IV$ data are related to the small dynamical critical exponent for frustrated JJA's.
As can be seen from the current dependences of the slope of the $IV$ curves, shown in Fig.\ \ref{fig2}, the $IV$ curves at high temperatures for all the $f$'s are Ohmic at low currents and bend upward at high currents.
At low temperatures, the $IV$ curves are bent downward.
The low-temperature curves can be fitted into a form $V \sim I \exp[-({I_T} /I)^{\mu}]$ with $\mu=$ 0.6$-$1.
The activated character of the low-temperature curves proposes a superconducting state as the low-temperature state.
The variation of qualitative nature of the $IV$ curves suggests that for all the $f$'s, a phase transition from a high-temperature resistive state to a low-temperature superconducting state occurs at the temperature where a straight $IV$ curve appears. 
The gradual drop of the resistance upon approaching the transition is indicative of a continuous phase transition.
For a continuous superconducting transition in 2D, $IV$ curves are expected to collapse onto two curves with $I$ and $V$ scaled by the general scaling form $V/I|T-T_{c}|^{z\nu} = {\cal E}_{\pm}(I/T|T-T_{c}|^{\nu})$, where $z$ is the dynamical critical exponent, $\nu$ the correlation-length critical exponent, and ${\cal E}_{\pm}$ the scaling functions above and below the transition temperature $T_c$.\cite{R15}
This scaling form becomes a simple power-law $IV$ relation $V \sim I^{z+1}$ at $T=T_c$ and $V/I \sim (T-T_{c})^{z \nu}$ in the low $I$ limit at $T > T_c$.
Figure \ref{fig3} shows the $IV$ curves scaled with the scaling form.
For all four $f$'s, we find the $IV$ data exhibit good scaling behaviors. 
A good scaling behavior is found for $f=$ 8/21 as well.
The scaling parameters $T_c$, $\nu$, and $z$ for $f=$ 8/21 are practically identical with those for $f=$ 0.382.
In the scaling analyses, we made use of the approximate values of $T_c$, $z$, and $\nu$ derived from the straight $IV$ curves in Fig.\ \ref{fig1} and the temperature vs resistance curves fitted into the relation $R \sim (T-T_{c})^{z\nu}$.
The values of $T_c$, $\nu$, and $z$ in the insets of Fig.\ \ref{fig3} are the refined ones in the scaling processes. 
The error bars, that is, the arbitrarinesses in determining the scaling parameters are 0.03 K for $T_c$, 0.05 for $z$, and 0.1 for $\nu$.
Distinct concavities of the $IV$ curves significantly limit the arbitrariness in determining the parameters.
The $IV$ data satisfy the recently proposed criterion \cite{R16} to determine if the data collapse is valid, as appeared in Fig.\ \ref{fig2}. 
The effect of finite-size-induced free vortices, which may alter the scaling behavior considerably in the absence of a magnetic field \cite{R17}, is not significant for our sample exposed to a magnetic field.\cite{R18}
We thus conclude that the scaling plots confirm for all the frustrations investigated a superconducting transition at the temperature where a straight $IV$ curve appears.

\begin{figure}[t]
\includegraphics[width=1.0\linewidth]{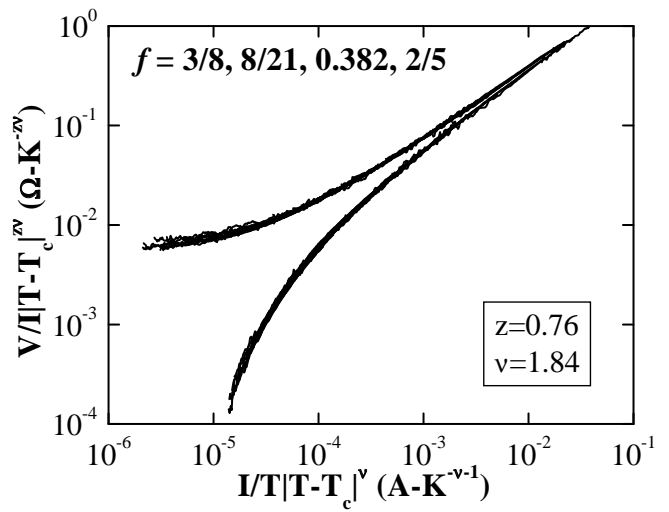}
\caption{Scaling plot of the $IV$ curves for $f=$ 3/8, 8/21, 0.382, and 2/5 with the same $z$ ($=0.76$) and $\nu$ ($=1.84$). The $IV$ data for four different frustrations collapse onto the nearly same curves.}
\label{fig4}
\end{figure}

The $T_c$'s for the $f$'s investigated correspond to (0.19-0.22)$J/k_{B}$, which is somewhat too high to be related with the vortex-pinning transition, the lower-temperature superconducting transition of two separate phase transitions observed in some MC simulations.\cite{R8} 
The vortex-pinning transition was expected to appear at $T_{c}<$ 0.1$J/k_{B}$ for $f=$ 3/8 and 0.382.
The weak $f$ dependence of $T_c$ does not seem to be compatible either with the argument \cite{R6} that $T_c$ should decrease to zero as $f$ approaches the irrational $f^*$.
The critical exponents $\nu$ and $z$ for $f=$ 3/8, 8/21, and 0.382 are identical within experimental errors with those of $f=$ 2/5.
In Fig.\ \ref{fig4}, we find that not only are the critical exponents similar for four different frustrations but the scaling functions are also.
However, the $IV$ data for $f=$ 5/12 which is only $\sim$ 0.017 distant from 2/5 fail to collapse onto the same curves in Fig.\ \ref{fig4}.
For $f=$ 1/3, it was found \cite{R19} that $\nu=$ 1.6 and $z=$ 0.60, significantly different from those of $f=$ 2/5.
Such approximately universal scaling behaviors for $f=$ 3/8, 8/21, 0.382, and 2/5 indicate that the superconducting transitions for $f=$ 3/8, 8/21, and 0.382 are similar in nature to that for the nearby simple frustration $f=$ 2/5.
For $f=$ 2/5, the superconducting transition has been understood in terms of the freezing of a vortex liquid into a pinned ordered vortex solid.\cite{R11,R20,R21}
The approximate universality also implies that the vortex configurations near $T_c$ for $f=$ 3/8, 8/21, 0.382, and 2/5 are quite similar to each other.
Adsorbed atoms on a periodic substrate which bear a strong resemblance to vortices in a periodic array of Josephson junctions may have a ground state near incommensuracy consisting of domains of nearby commensurate phase(s) separated by domain walls.\cite{R22}
If the JJA has domains of the $f=$ 2/5 phase separated by quasiperiodic or commensurate domain walls as the ground states for $f=$ 3/8, 8/21, and 0.382 analogous to the adsorbate systems near incommensurability, the vortex configurations for $f=$ 3/8, 8/21, 0.382, and 2/5 may look similar in the vicinity of transition where many extra domain walls get excited.
Extending our finding at the $f$'s around 2/5, one might expect that the behaviors of vortices in the wide range of frustrations could be understood from those of a few simple rational frustrations at which the system has distinct local minima in energy.
The $IV$ characteristics for $f=$ 5/12 might then be understood from those of possibly $f=$ 3/7.

As discussed above, our experimental finding is inconsistent with the numerical results indicating either a glasslike transition or two separate transitions at the irrational $f^*$ or its nearby high-order rational $f$'s.
Instead, it seems to be compatible to some extent with the simulation results \cite{R9} indicating for a sequence of $f$'s approaching the irrational $f^*$ a single finite-temperature ordering transition of a vortex liquid into a striped lattice phase with domains of the $f=$ 8/21 staircase state separated by parallel walls.
Denniston and Tang \cite{R9} argued that the conflicting simulation results near the irrational $f^*$ are due to the impositions of different boundary conditions.
An imposition of an improper boundary condition incompatible with incommensurate or long-period commensurate phases may produce unrealistic simulation results near incommensurability.
Another possible cause for the inconsistencies between the experiments and the simulations is the disorder effects.
A real array always contains some amount of quenched disorder, for instance, the inevitable variation of the junction coupling strength.
Even a small amount of quenched disorder has been found \cite{R11,R12,R21,R23,R24} to have a significant effect on phase transitions of JJA's at simple rational frustrations.
Therefore, for the present, we may not totally exclude the disorder effects as a possible cause for the inconsistencies.

In conclusion, approximately universal scaling behaviors of the $IV$ characteristics are found for a square JJA at $f=$ 3/8, 8/21, 0.382 [$\approx (3-\sqrt{5})/2$], and 2/5.
The approximately universal scaling behaviors and high superconducting transition temperatures indicate that both the nature of the superconducting transition and the vortex configuration near the transition at the high-order rational frustrations $f=$ 3/8, 8/21, and 0.382 are similar to those at the nearby simple frustration $f=$ 2/5.
This finding suggests that the behaviors of Josephson junction arrays in the wide range of frustrations might be understood from those of a few simple rational frustrations.

This work was supported by the Ministry of Education through the BK21 program.


\begin{references}

\bibitem {R1}
See, for example, J. P. Straley and G. M. Barnett, Phys. Rev. B {\bf {48}}, 3309 (1993).

\bibitem {R2}
S. Teitel and C. Jayaprakash, Phys. Rev. Lett. {\bf {51}}, 1999 (1983);
T. C. Halsey, J. Phys. C {\bf {18}}, 2437 (1985).

\bibitem {R3}
M. Franz and S. Teitel, Phys. Rev. B {\bf {51}}, 6551 (1995);
S. A. Hattel and J. M. Wheatley, {\it{ibid}}. {\bf{51}}, 11 951 (1995).

\bibitem {R4}
T. C. Halsey, Phys. Rev. Lett. {\bf {55}}, 1018 (1985).

\bibitem {R5}
E. Granato, Phys. Rev. B {\bf{54}}, R9655 (1996).

\bibitem {R6}
S. Y. Park, M. Y. Choi, B. J. Kim, G. S. Jeon, and J. S. Chung, Phys. Rev. Lett. {\bf{85}}, 3484 (2000);
M. Y. Choi, J. S. Chung, D. Stroud, and J. Choi, Phys. Rev. B {\bf {40}}, 5147 (1989).

\bibitem {R7}
M. P. A. Fisher, Phys. Rev. Lett. {\bf{62}}, 1415 (1989);
C. Dekker, P. J. M. W$\ddot{\rm{o}}$ltgens, R. H. Koch, B. W. Hussey, and A. Gupta, {\it{ibid}}. {\bf{69}}, 2717 (1992);
R. A. Hyman, M. Wallin, M. P. A. Fisher, S. M. Girvin, and A. P. Young, Phys. Rev. B {\bf{51}}, 15 304 (1995).

\bibitem {R8}
P. Gupta, S. Teitel, and M. J. P. Gingras, Phys. Rev. Lett. {\bf {80}}, 105 (1998);
S. J. Lee, J.-R. Lee, and B. Kim, {\it {ibid}}. {\bf {88}}, 025701 (2002);
M. R. Kolahchi and H. Fazli, Phys. Rev. B {\bf {62}}, 9089 (2000).

\bibitem {R9}
C. Denniston and C. Tang, Phys. Rev. B {\bf {60}}, 3163 (1999).

\bibitem {R10}
F. Yu, N. E. Israeloff, A. M. Goldman, and R. Bojko, Phys. Rev. Lett. {\bf {68}}, 2535 (1992).

\bibitem {R11}
X. S. Ling, H. J. Lezec, M. J. Higins, J. S. Tsai, J. Fujita, H. Numata, Y. Nakamura, Y. Ochiai, Chao Tang, P. M. Chaikin, and S. Bhattacharya, Phys. Rev. Lett. {\bf {76}}, 2989 (1996).

\bibitem {R12}
Young-Je Yun, In-Cheol Baek, and Mu-Yong Choi, Phys. Rev. Lett. {\bf {89}}, 037004 (2002).

\bibitem {R13}
The positions of resistance minima at simple rational $f$'s could be identified in the magnetoresistance curve with uncertainty $\Delta f \le 0.0005$.
The calculated magnetic-field inhomogeneity over the sample due to the finite length of the solenoid is $\sim 0.05$\%.
The magnetic-field inhomogeneity due to the presence of trapped vortices in the sample was $\le 0.1$\%, estimated from the amount of hysteresis in the magnetoresistance curve appearing when the field-sweeping direction was reversed.

\bibitem {R14}
P. G. de Gennes, Rev. Mod. Phys. {\bf {36}}, 225 (1964).

\bibitem {R15}
D. S. Fisher, M. P. A. Fisher, and D. A. Huse, Phys. Rev. B {\bf {43}}, 130 (1991).

\bibitem {R16}
D. R. Strachan, M. C. Sullivan, P. Fournier, S. P. Pai, T. Venkatesan, and C. J. Lobb, Phys. Rev. Lett. {\bf {87}}, 067007 (2001).

\bibitem {R17}
J. Holzer, R. S. Newrock, C. J. Lobb, T. Aouaroun, and S. T. Herbert, Phys. Rev. B {\bf {63}}, 184508 (2001);
K. Medvedyeva, B. J. Kim, and P. Minnhagen, {\it {ibid}}. {\bf {62}}, 14 531 (2000);
S. W. Pierson, M. Friesen, S. M. Ammirata, J. C. Hunnicutt, and L. A. Gorham, {\it {ibid}}.  {\bf {60}}, 1309 (1999).

\bibitem {R18}
For our sample with $I=10^{-5}$ A at $T=6.0$ K, $V(f=0.382) \approx 2.2 \times 10^{-8}$ V, whereas $V(f=0) < 10^{-10}$ V, a part of which is due to the finite-size-induced vortices.

\bibitem {R19}
The scaling plot of the $IV$ data for $f=$ 1/3 is shown in Ref. \cite{R12}.
The $IV$ measurements were performed on the same sample used for the present work.

\bibitem {R20}
Y.-H. Li and S. Teitel, Phys. Rev. Lett. {\bf {65}}, 2595 (1990).

\bibitem {R21}
C. Denniston and C. Tang, Phys. Rev. Lett. {\bf {79}}, 451 (1997).

\bibitem {R22}
See, for example, P. M. Chaikin and T. C. Lubensky, {\it {Principles of Condensed Matter Physics}} (Cambridge University Press, New York, 1995), pp. 601-620.

\bibitem {R23}
J. M. Kosterlitz and M. V. Simkin, Phys. Rev. Lett. {\bf {79}}, 1098 (1997);
P. Gupta and S. Teitel, {\it {ibid}}. {\bf {82}}, 5313 (1999), and references therein.

\bibitem {R24}
M. Benakli, E. Granato, S. R. Shenoy, and M. Gabay, Phys. Rev. B {\bf {57}}, 10 314 (1998);
E. Granato and D. Dom\'{\i}nguez, {\it {ibid}}. {\bf {63}}, 094507 (2001).
\end{references}
\end{document}